\documentclass[10pt,twocolumn,letterpaper]{article}

\usepackage{ijcb}
\usepackage{times}
\usepackage{epsfig}
\usepackage{graphicx}
\usepackage{amsmath}
\usepackage{amssymb}
\usepackage[norule,symbol,perpage]{footmisc}
\usepackage{caption}
\usepackage{subcaption}
\usepackage{textcomp}
\usepackage{comment}
\usepackage{xcolor}
\usepackage[inline]{enumitem}
\usepackage{float}
\usepackage{algorithmic}
\usepackage{lipsum}
\usepackage{fancyhdr}
\usepackage{flushend}

\usepackage[pagebackref=true,breaklinks=true,colorlinks,bookmarks=false]{hyperref}

\ijcbfinalcopy 

\ifijcbfinal\fi

\makeatletter
\def\ps@IEEEtitlepagestyle{
\def\@oddfoot{\mycopyrightnotice}
\def\@evenfoot{}
}
\def\mycopyrightnotice{
{\hfill \footnotesize 978-1-7281-9186-7/20/\$31.00 \copyright 2020 IEEE\hfill}
}
\makeatother

\begin{document}

\title{Analysis of Dilation in Children and its Impact on Iris Recognition}

\author{Priyanka Das, Laura Holsopple, Stephanie Schuckers\\
Clarkson University\\
Potsdam, NY, USA\\
{\tt\small prdas@clarkson.edu,lholsopp@clarkson.edu,sschucke@clarkson.edu}
\and
Michael Schuckers\\
St. Lawrence University\\
Canton, NY, USA\\
{\tt\small schuckers@stlawu.edu }
}

\maketitle
\thispagestyle{empty}

\begin{abstract}
The dilation of the pupil and it's variation between a mated pair of irides has been found to be an important factor in the performance of iris recognition systems. Studies on adult irides indicated significant impact of dilation on iris recognition performance at different ages. However, the results of adults may not necessarily translate to children. This study analyzes dilation as a factor of age and over time in children, from data collected from same 209 subjects in the age group of four to 11 years at enrollment, longitudinally over three years spaced by six months. The performance of iris recognition is also analyzed in presence of dilation variation.  
\end{abstract}

\let\thefootnote\relax\footnotetext{\mycopyrightnotice}

\section{Introduction}
Iris recognition is a well established field within biometric recognition. In the last decade, efforts were targeted towards studying performance in systems longitudinally, i.e. as the time between the enrollment image and subsequent probe image increases. Dilation has been identified as an important factor contributing to the variability in the iris recognition performance in multiple studies \cite{ortiz2013linear} \cite{hollingsworth2009pupil} \cite{fairhurst2011analysis}.  \par 
Dilation is defined as the degree to which the pupil is dilated or constricted and has been represented as a dimensionless quantity of the pupil to iris ratio \cite{iris_standard_report}. A high ratio indicates a high degree of pupil dilation and thus a low usable iris area for analysis. A lower ratio indicates a constricted pupil that could complicate iris segmentation, impacting performance adversely. Intra-subject pupil size  may vary due to physiological factors like age, aging, in response to emotional stimuli and environmental factors like illumination and medical conditions and can vary stochastically over a time frame of few seconds to decades. Age, aging and illumination are quantifiable factors. In a practical scenario of iris recognition, illumination and environmental factors may vary. Identifying and quantifying the impact of age and aging on dilation and iris recognition performance could help improve the robustness of the existing iris recognition technologies. Multiple studies reported the impact of age on dilation and iris recognition performance in adults. However, the age group of children between 0 to 18 years remains unstudied. As indicated in a previous study \cite{adler1965physiology}, as children grow, they have a growth factor which impacts the size of the pupil from birth to adolescence. Results from studies involving adults may not translate to children.\par
This study looks into the \textit{impact of age and aging on dilation} and the \textit{impact of variation in dilation on the iris recognition performance in children} in the age group of four to 11 years at enrollment from 209 subjects over a period of three years.

\section{State of Art : Dilation and Iris Recognition}

 To understand the effect of dilation on iris recognition performance, it is important to understand how iris recognition operates \cite{daugman2009iris}. For the purpose of biometric recognition, the pupillary boundary and the limbus boundary are detected from the iris image and are segmented. The annular region representing the iris is projected into a pseudo-polar coordinate system drawing analogy to a homogeneous rubber sheet model. The elastic mesh work of the iris causing the dilation and constriction is modelled by this coordinate system by drawing analogy from the topology of a homogeneous rubber sheet annulus anchored along its outer perimeter with the tension controlled by an interior ring of variable radius \cite{daugman2009iris}. The coordinate system has a polar variable i.e. the angle, $\theta$, ranging between 0 and 2$\pi$ and a radial variable, r, of the annular region which is always an unit interval [0,1], both being dimensionless. Each point of the annular region i.e. the iris, irrespective of dilation, is represented by a pair of variables [r,$\theta$]. Since the radial variable ranges from the pupillary boundary to the limbus as a unit interval, it inherently corrects the deformation in the iris due to variable pupil dilation. This allows approximated comparison between irides with variation in dilation, introducing non-affine pattern deformation. However, this model considers the deformation linear. There are studies affirming non-linear deformation pattern as well \cite{wyatt2000minimum}. Thus, the rubber sheet model, though widely used, does not always provides an absolute representation of the iris having deformity as a function of pupil dilation. This has reflected in various studies with poor biometric identification performance in cases with high degree of dilation variation, even to a point of false rejection during biometric identification \cite{ortiz2013linear} \cite{hollingsworth2009pupil} \cite{fairhurst2011analysis}. Over three decades even though Daughman's rubber sheet retained its popularity, multiple alternatives to the model has been published in literature \cite{wildes1996machine} \cite{tisse2002person} \cite{ma2002iris} \cite{ma2004efficient} \cite{monro2007dct} \cite{sun2008ordinal} \cite{daugman2009iris} \cite{liu2016deepiris} \cite{gangwar2016deepirisnet} \cite{nguyen2017iris} \cite{arsalan2017deep}. Most commercial algorithms however, remains a blackbox with no public information on the techniques adopted in their algorithm. Thus, the results of matching pair of iris images with varying dilation may be impacted differently with different algorithms.\par
Pupil dilation or constriction is governed by the dilator and sphincter muscles in the iris which are controlled by the sympathetic and parasympathetic nervous system, respectively. The reflex action of dilation or constriction is effected by various variable factors, including illumination, emotional and non-emotional factors and medical conditions. Physiological factors of age and aging is also related to dilation. Fluctuation in pupil dilation is highly correlated with emotion processing and non-emotional state of decision making \cite{oliva2018pupil}. Many medical conditions may affect the pupil dilation. For opthalmological examination a chemical compound atropine sulphate is used to dilate the pupil which remains effective for several hours. In 1950, Birren et al. \cite{birren1950age} reported significant reduction in pupil size with age in both light and dark  conditions, from a study on 222 subjects in the age group of 20 to 89 years of age. In 1965 Alder reported from his medical research that pupils are small in newborn babies and remain small until the first year of birth, reaching its maximum during childhood and adolescence, and then gradually decreases with advancing age \cite{adler1965physiology}. More recently, in 1994 Winn et.al. investigated pupil dilation at fixed illumination levels in 91 subjects in the age group of 17 and 83 years  and concluded a steady linear decay in the dilation as a function of age for all illuminance levels \cite{winn1994factors}. In 2008 Hollingsworth et.al. \cite{hollingsworth2009pupil} studied the performance of iris recognition with 18 adult subjects and noted that, as the size of the pupil increases, the mean of the hamming distance gets closer to the non-match distribution. The researchers also noted a larger difference in pupil size between enrollment and verification yields higher dissimilarity. In 2011, Fairhurst and Erbilek  \cite{fairhurst2011analysis} in a study with 632 images from 79 subjects in the age group of 18-73 years concludes a gradual decay in mean dilation from the age group $<$25, to 25-60 to $>$60 years. Aging affects the accommodation capacity of the pupil, resulting in decreased dilation with aging, keeping other factors constant. The researchers concluded that, in older adults (age $>$ 60) the iris recognition performance is less impacted by dilation as the change in dilation is decreased in this age group as opposed to that of `younger individuals' (their study only included adults). An increase in performance is noted with older age group with state of art segmentation. In 2013, Ortiz et al. \cite{ortiz2013linear} studied effect of age and aging on biometric performance using data from 955 subjects in the age group of 18 to 64 years collected over 3 years. The age group between 18 and 25 years dominated the subject count. They observed an increase in the hamming distance between mated pair of images with age and hypothesized that the increased hamming distance is correlated with age as change in the size of the pupil is impacted by age. \par
It is important to note that the `effect of age' and `effect of aging' on any trait are two different concepts as defined below-
\begin{itemize}[leftmargin=*, noitemsep]
\item \textbf{Age study :}\label{CSstudy}\par
Impact of age of an individual on dilation is investigated in this study by maximizing the use of our longitudinal dataset. We re-organized the data by age regardless of what session they were captured, where each age between four to 14 years is a cohort and performed a basic analysis of all data across age to understand dilation at different ages. 

\item \textbf{Aging study or Longitudinal study :}\label{Lstudy}\par
Aging effects are typically investigated by comparing the performance of a sample or samples at different points in time i.e. longitudinal study \cite{corso1971sensory}. 
\end{itemize}
Our literature review revealed that most prior researches were cross-sectional studies on the relationship between dilation and age, predominantly on large groups of adult subjects and conclusions were drawn statistically on widely spanned age groups. Only one research group, Ortiz et al. \cite{ortiz2013linear} correlated aging, dilation and iris recognition performance (hamming distance and match score) by designing a composite model and concluded a measurable degradation in the performance metric due to dilation difference caused by aging effect. We did not find any literature on the effect of pupil size variation on iris recognition performance in children. This work concentrates on quantitative analysis of dilation as a factor of both age and aging and its effect on biometric performance in children in the age group of 4 to 14 years at a granular level for each age, from 8612 samples collected from the same 209 unique subjects over three years from seven collection sessions spaced by approximately six months. This paper addresses mainly on the three following questions for ages four to 14 years:
\begin{itemize}[leftmargin=*, itemsep=0.1ex]
    \item \textbf{Is there a relationship between age and dilation?}
    \item \textbf{Does aging impact dilation over a period of three years?}
    \item \textbf{How does the difference in dilation between mated pair of images impact iris recognition performance in a longitudinal scenario of three years?}  
\end{itemize}

\section{ Definitions and Acronyms} \label{Def}
This section summarizes the terms and the formulas used throughout this paper.
\subsection{Dilation}\label{Dil_def}
Dilation or pupil dilation is a dimensionless quantity measuring the degree to which the pupil is dilated or constricted, measured as a ratio of pupil radius and iris radius as defined below.
\begin{equation}\label{eq:D}
 Dilation(D) = \frac{Pupil\: radius}{Iris\: radius} \times 100
\end{equation}

\subsection{Delta Dilation}
Difference in the pupil dilation between mated pair of iris images is defined as Delta Dilation ($\Delta$D) in this paper and the measure follows NIST work in \cite{grother2013irex} as below.
\begin{equation}
    Delta\:Dilation (\Delta D) = 1 - \frac{1-\frac{D1}{100}}{1-\frac{D2}{100}}
\end{equation}
considering, D1 $\geq$ D2, where, D1 and D2 are the pupil dilation of the first and the second iris images as estimated using equation \ref{eq:D}.

\subsection{Abbreviations used in the paper}\label{Vocab}

\begin{enumerate}[leftmargin=*, noitemsep]
   \item\textbf{RI}: Right Iris
  \item \textbf{LI}: Left Iris
  \item \textbf{MS}: Match Score
  \item \textbf{G1}: Group 1 - All subjects who participated in \textbf{at least} two of the seven sessions
  \item \textbf{G2}: Group 2 - All subjects who participated in Collection 1 \textbf{and} Collection 7 with possible intermittent gap
  \item \textbf{G3}: Group 3 - All subjects who participated in \textbf{all} seven sessions from Collection 1 \textbf{to} Collection 7
\end{enumerate}

\section{Data Collection Protocol and Statistics}

The iris data used in this study is part of a larger longitudinal dataset of multiple biometric modalities collected from the same children over three years. This is a continuing study and till the point of analysis and preparation of this paper we had in total seven sessions of data collection. Data was collected from 239 subjects in the age group of four years to 11 years at the time of enrollment with seven visits subsequently spanned over three years, spaced by approximately six months. 209 subjects participated in more than one session, and data from these subjects were analyzed.\par
Researchers collaborate with the local school to identify subjects for voluntary participation. An approved IRB protocol requires an informed consent from parents and participants. Initial participation for the first session was open to children aged between 4 to 11 years. Henceforth, every year new subjects from Pre-K are added to the study who are mostly in the age group of four to five years. The equipment are setup in the school in an isolated room as provided by the school for the entire collection week(s). The same equipment are used for each session to minimize variation in the data quality or properties. However, the room may vary at each session based on availability, which might affect the collection environment like lighting, temperature, noise which may impact the collected biometric data. Measurements are taken to mitigate the variation in environmental factors effecting iris data collection. The blinds in the room are drawn to minimize exposure from external daylight and NIR from the sun which is the primary illumination for iris capture. Factors such as medical conditions, different collection rooms, weather conditions, time of year (fall or spring) cannot be eradicated. However, we provide each subject considerable time in the collection room to allow the eye to optimize and accommodate and adjust the dilation, to mitigate the variability due to variation in illumination before coming for collection. Since the data collections are done indoor we expect negligible impact due to weather and season. Participation on the day of the collection is voluntary; if a subject refuses to provide data on the day of the collection they are excused from participating. Participants are provided with an amicable environment, with no emotional excitation/stimulating factor which might impact the data collection. However, any personal emotional factor is not accounted in the data. A commercially available iris sensor, IG-AD100 Dual Iris Camera manufactured by Iris Guard is used for data capture which detects and auto-captures iris in the NIR wavelength. The quality of the camera and the images captured are compliant with ISO/IEC 19794-6 \cite{iris_standard_report}. In addition to the NIR illumination, the camera also has a white LED flashing illumination with the purpose of providing stimulus for stabilizing the pupil dilation to maintain a stable intra-subject dilation across sessions. The camera was donated for the purpose of the study of the impact on iris recognition in children.\par
\begin{figure}[t]
\centering
\includegraphics[width=1\linewidth]{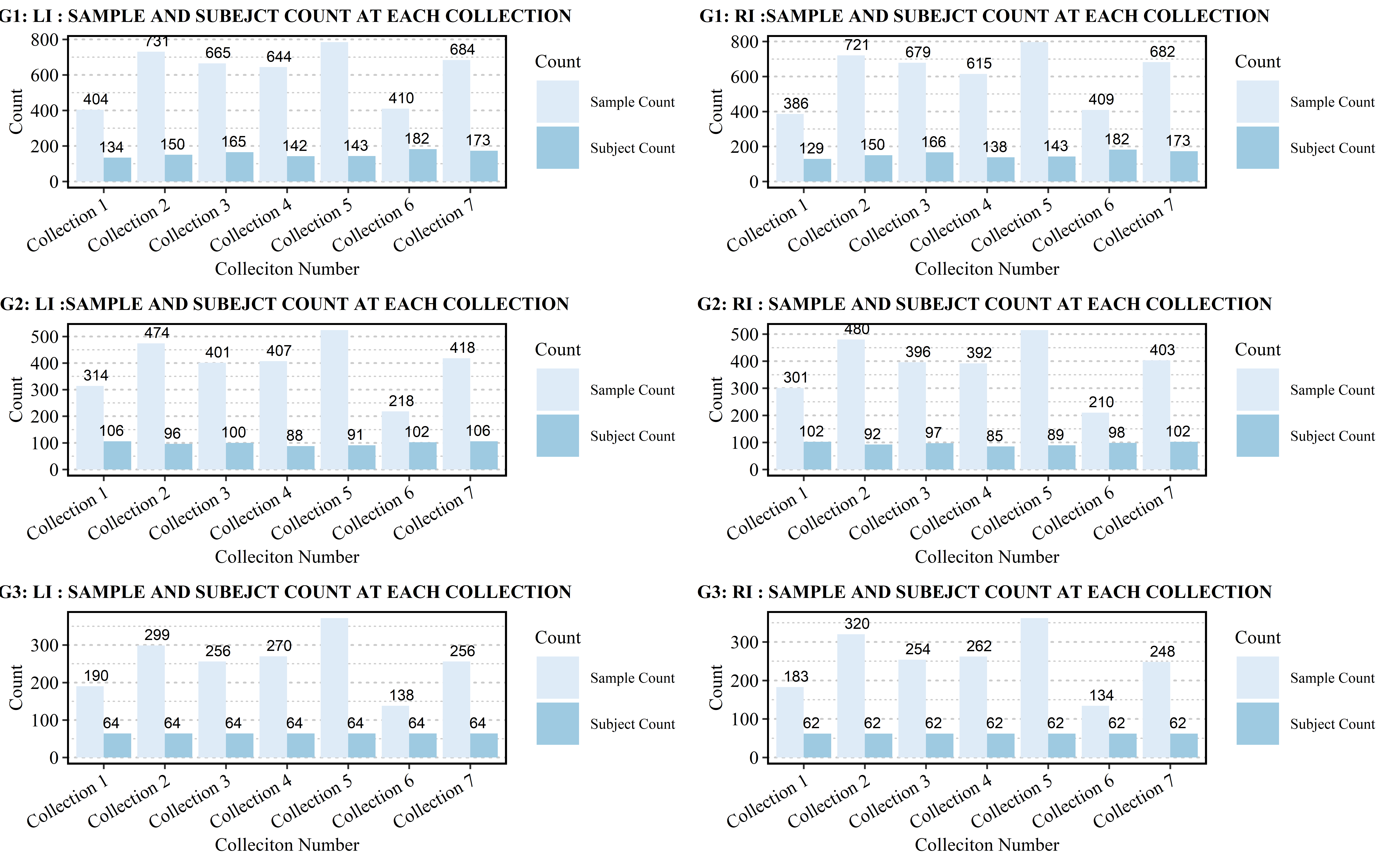}

    \caption{Number of subjects and image samples at different collection for different groups - G1, G2 and G3 (top to bottom) of study for both left and right irides (left and right, respectively)}
    \label{fig:Sample_sub_count}
\end{figure}

Four images were captured from each eye at each session with some exceptions in first session and in sixth session when at least two images were captured. A few subjects may have more than four images captured per eye per session. Prior to the seventh session, the images collected were highly correlated, due to the internal setup of the camera which captures images within a few seconds. The protocol was modified in the seventh session when the four images were captured in two sets with a small time gap (less than two minutes) between the sets. A total of 8612 samples (LI: 4323, RI: 4289) were analyzed. Right and left iris images were analyzed separately. The number of participants in each session may vary due to new subjects added to the study every year, subjects moving out of the study and absentees on the day of the collection. The data has been studied in three groups, G1, G2 and G3, as described in Section \ref{Vocab}. Subject and sample count in each visit for each group is shown in Figure \ref{fig:Sample_sub_count}. 209 unique subjects have participated in more than one collection. Thus in G1- 209 (209 for RI) , in G2 - 105 (101 for RI)  and in G3 - 63 (62 for RI) unique subjects were analyzed. The discrepancy in the number of subjects between LI and RI is due to the fact that for a few subjects both the irides could not be captured.

\section{Results}
The impact of age, ranging between four to 14 years, on dilation and longitudinal impact of aging over three years for the age group of four to 11 years on delta dilation ($\Delta$D) has been analyzed in this section. Further, the impact of $\Delta$D on longitudinal performance of match score has been analyzed. All analysis is done based on groups (G1, G2, G3) as defined in Section~\ref{Vocab}. All attributes (pupil and iris radius) has been extracted and matching has been performed with commercially available software Verieye v11.1. Verieye complies by ISO/IEC 19794-6 \cite{iris_standard_report} guidelines in processing images. It is important to use a commercial ISO standardized software to study aging. All further analysis has been performed in MATLAB 2018b and R studio v1.1.456.

\subsection{Age as a factor of dilation}\label{A_F_D}

\begin{figure*}[t]
\centering
\begin{subfigure}{0.45\textwidth}
  \centering
  \includegraphics[width=1\linewidth, height= 4.2 cm]{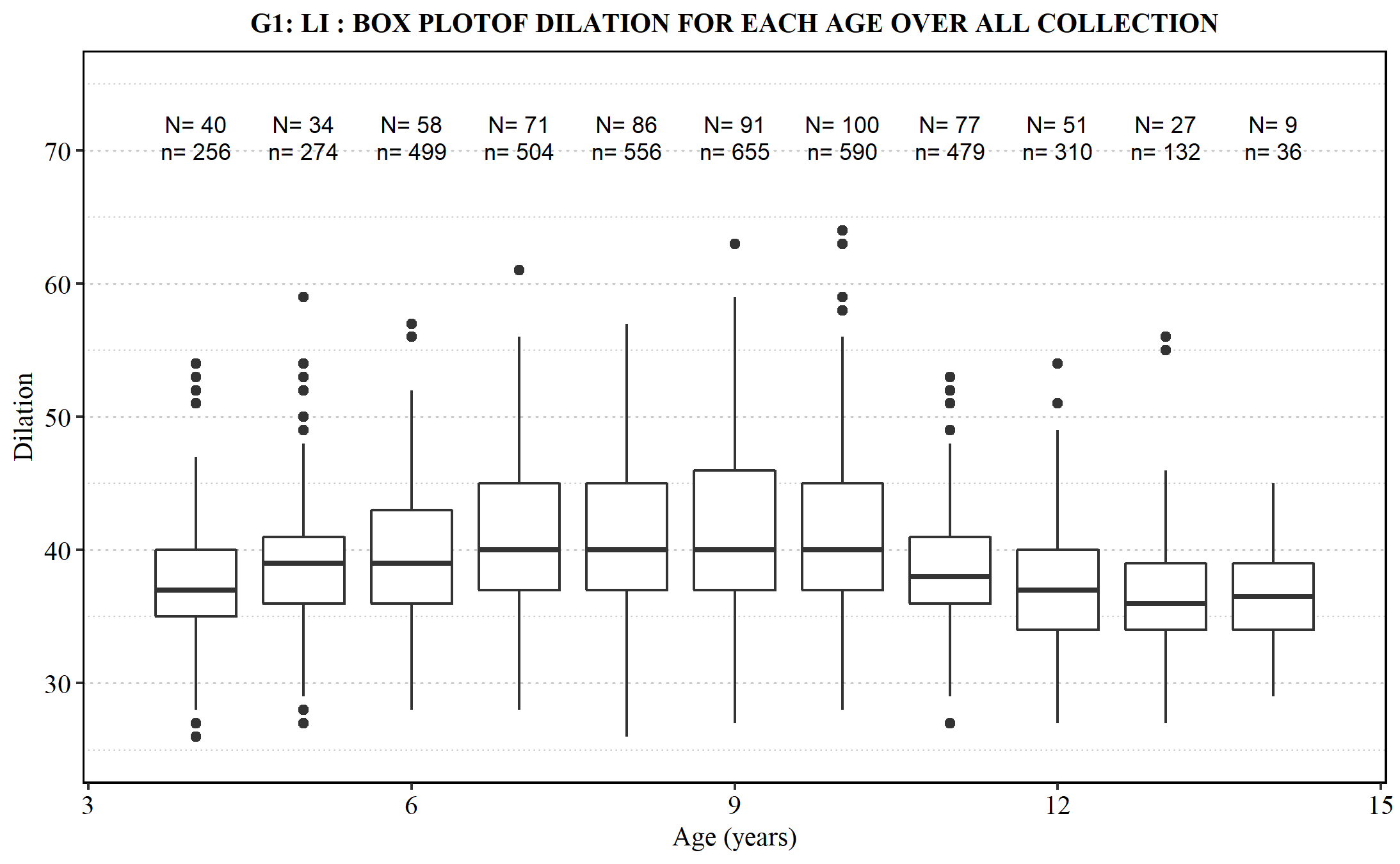}
  \label{fig:G1_L_age_f_Dil}
\end{subfigure}
~
\begin{subfigure}{0.45\textwidth}
  \centering
  \includegraphics[width=1\linewidth, height= 4.2 cm]{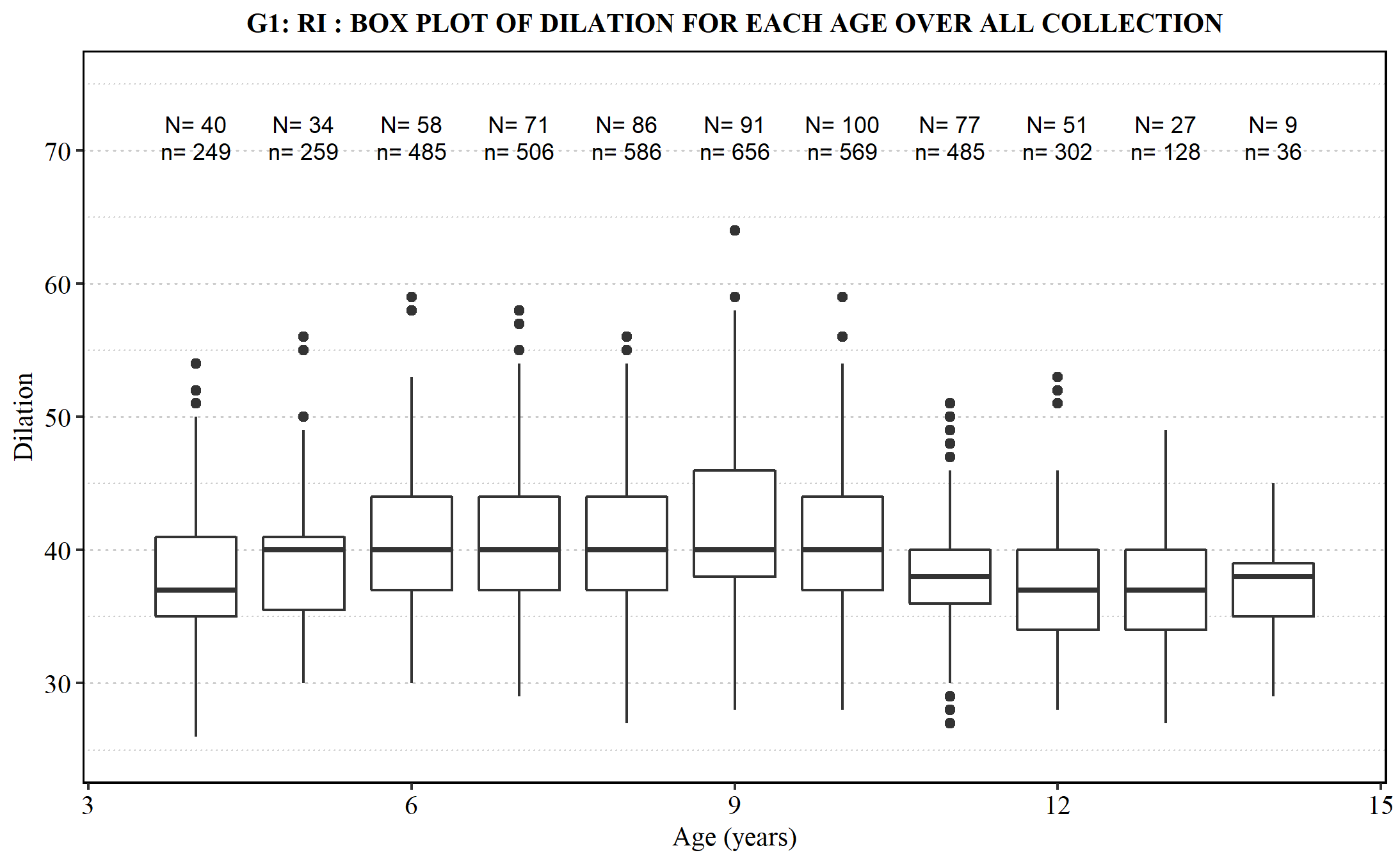}
  \label{fig:G1_R_age_f_Dil}
\end{subfigure}

\hfill

\begin{subfigure}{0.45\textwidth}
  \centering
  \includegraphics[width=1\linewidth, height = 4.2 cm]{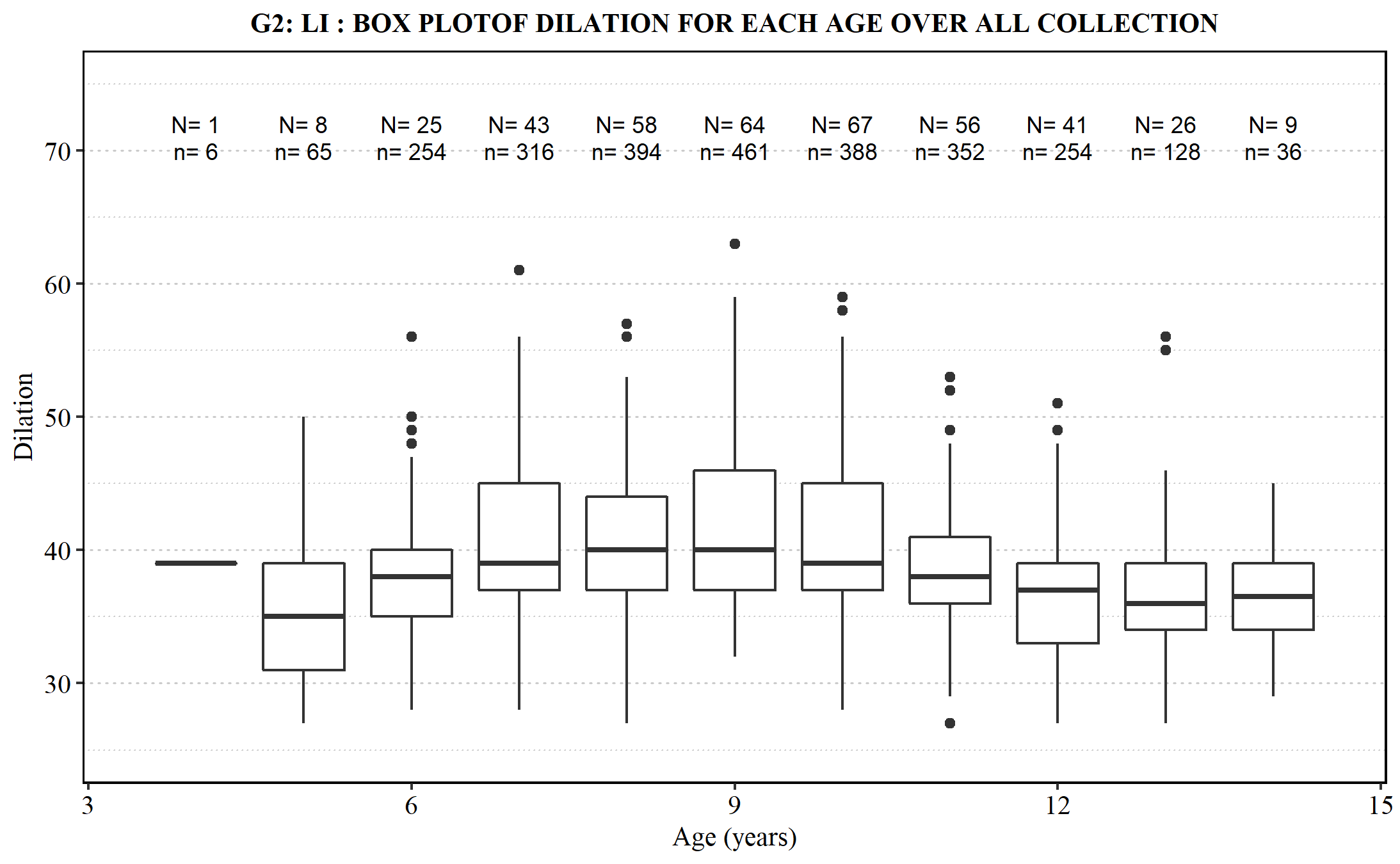}
  
  \label{fig:G2_L_age_f_Dil}
\end{subfigure}
~
\begin{subfigure}{0.45\textwidth}
  \centering
  \includegraphics[width=1\linewidth, height = 4.2 cm]{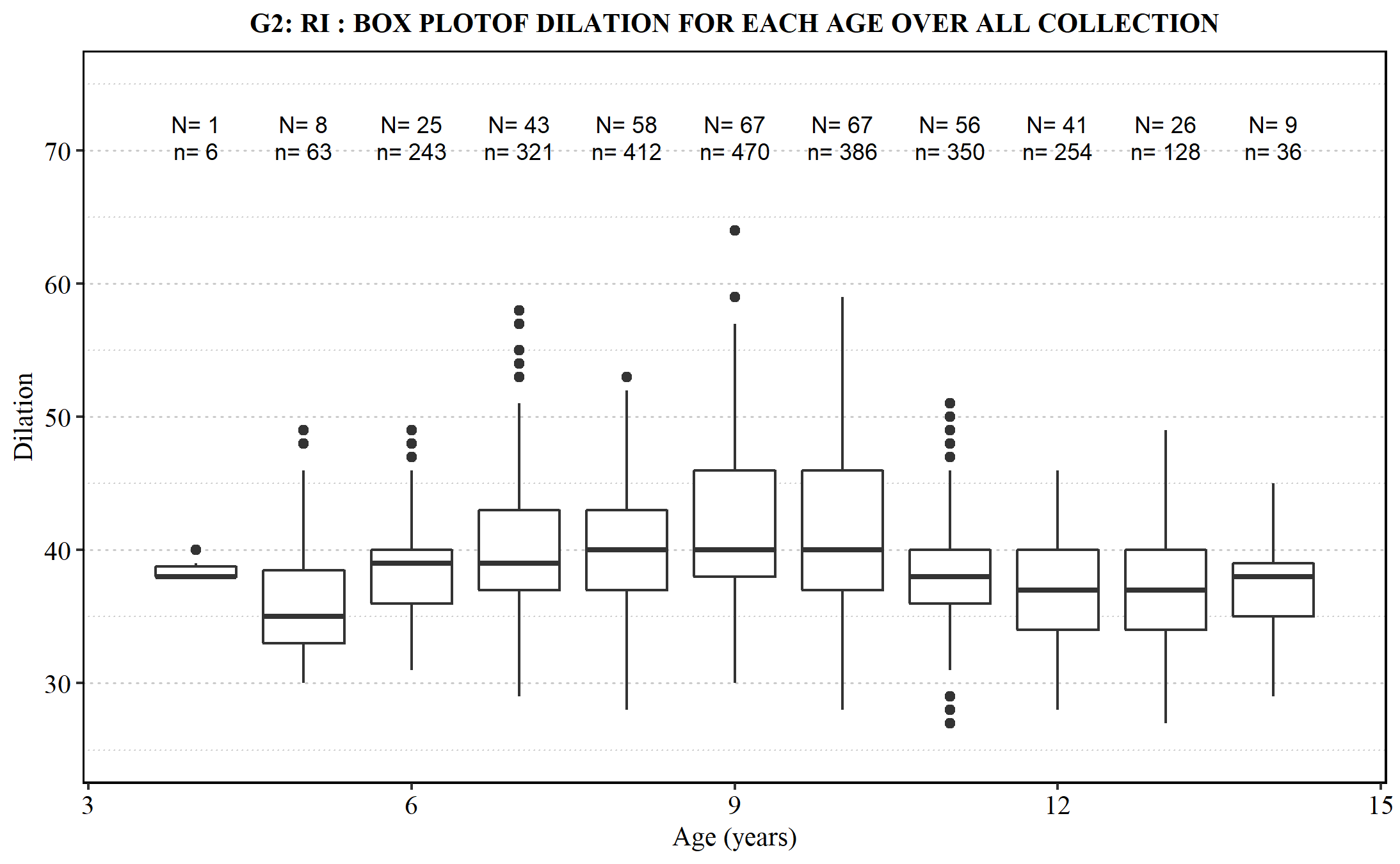}
  
  \label{fig:G2_R_age_f_Dil}
\end{subfigure}

\hfill

\begin{subfigure}{0.45\textwidth}
  \centering
  \includegraphics[width=1\linewidth, height = 4.2 cm]{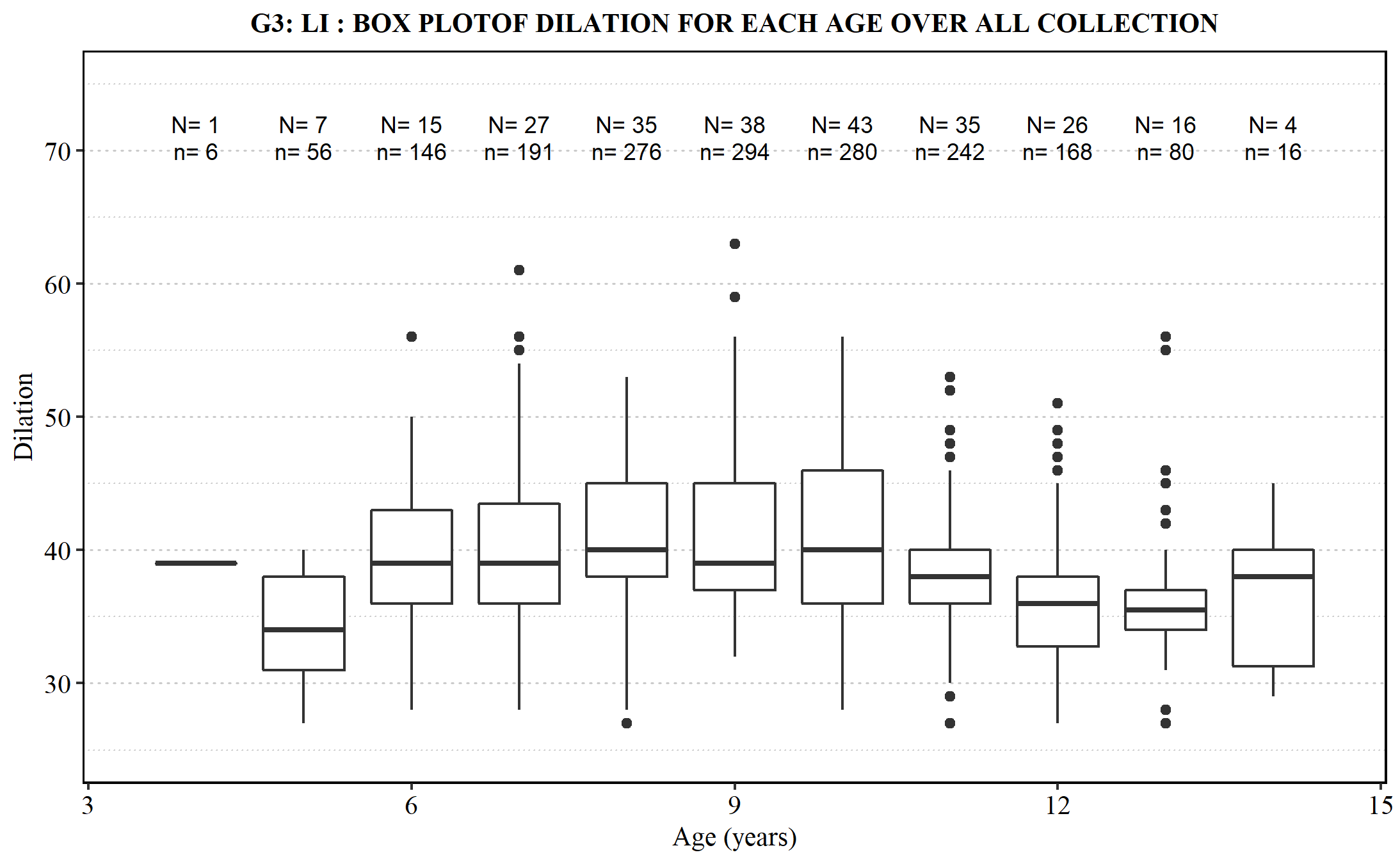}
  \label{fig:G3_L_age_f_Dil}
\end{subfigure}
~
\begin{subfigure}{0.45\textwidth}
  \centering
  \includegraphics[width=1\linewidth, height = 4.2 cm]{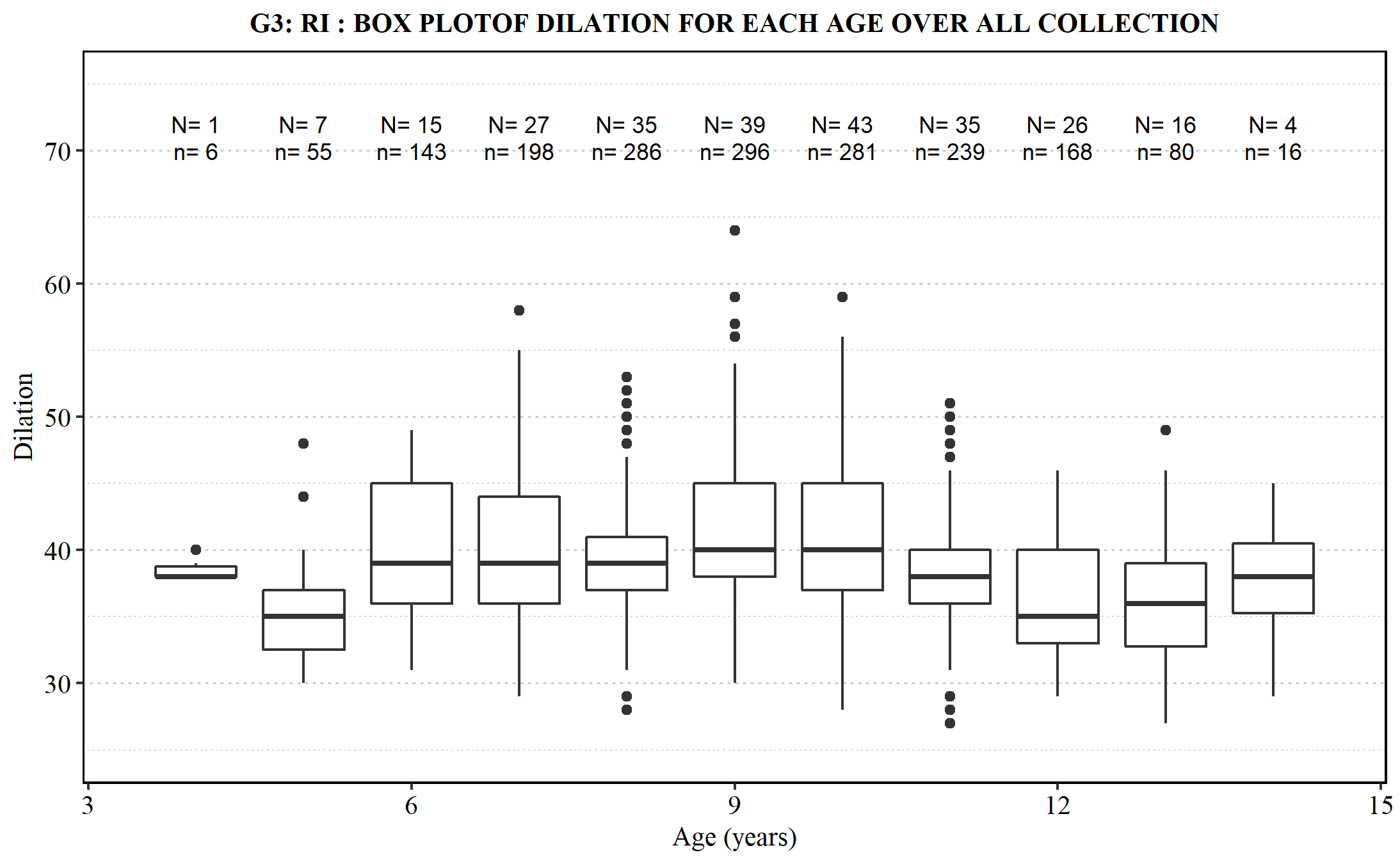}
\label{fig:G3_R_age_f_Dil}
\end{subfigure}

\caption{Boxplot of dilation for different age groups for all three groups of study - G1, G2, and G3 (top to bottom, for both LI and RI(left and right, respectively). 'N' denotes subject count and 'n' denotes sample count for each boxplot.}
\label{fig:Age_Dil}
\end{figure*}

To analyze the correlation between age and dilation, each age between four to 14 years is considered as a cohort; images in the dataset from each age are analyzed for each cohort irrespective of the session the data was captured. Figure~\ref{fig:Age_Dil} shows the dilation distribution in boxplots for each age cohort for all three groups of study (G1, G2, G3) for both LI and RI and respective sample and subject count.\par
For G1, data from a significant number of subjects are available for each age between four to 13 years. While the mean does not appear to vary across ages, there is increased variability in ages 7-10 years compared to 4-6 and 11-14 years, particularly between 50th percentile and the maximum value of dilation, i.e. there was more spread for those subjects in the upper half of dilation. Said simply, the upper half of subjects in ages 7-10 had a higher dilation compared to other ages. The 75th percentile changes from around 40 for ages 4-6, to 45 for ages 7-10, and finally comes back around 40 for ages above 10. Outliers are observed at almost all ages. The mean dilation shows a partial pattern across ages- being minimum at age four and gradually increasing till the age group of 8-10 years and again gradually decreasing or plateauing till the age of 14 years, particularly prominent in G2 and G3. This pattern partially corroborates the observation by Alder in 1965 in \cite{adler1965physiology} that the pupil reach its maximum size between first year of birth to childhood and adolescence, and then gradually decreases with advancing age. While this trend is not clearly evident in younger ages in this dataset given the large variability, there appears to be a downward trend after age 10. The trends observed for G1 are similar for both G2 and G3. It should also be noted that the subject count for age 4,5 and 14 years in G3 is low and is not likely to produce reliable statistical plot.

\subsection{Aging as a Factor of Dilation}
\begin{figure*}[!ht]
\centering
\begin{subfigure}{1\textwidth}
  \centering
  \includegraphics[width=0.8\linewidth, height= 7 cm]{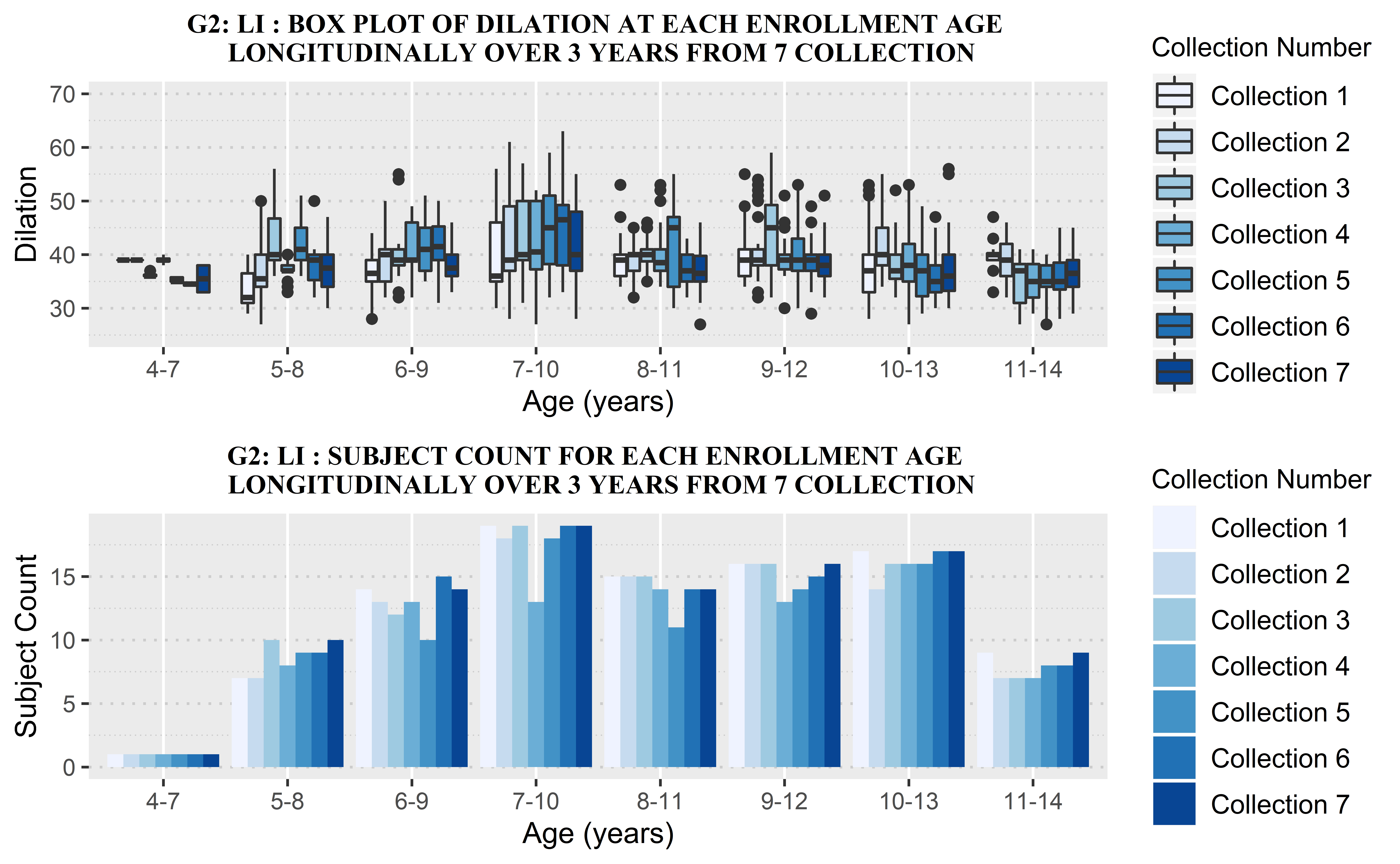}
  \label{fig:G2_L_aging_f_Dil}
\end{subfigure}

\begin{subfigure}{1\textwidth}
  \centering
  \includegraphics[width=0.8\linewidth, height= 7 cm]{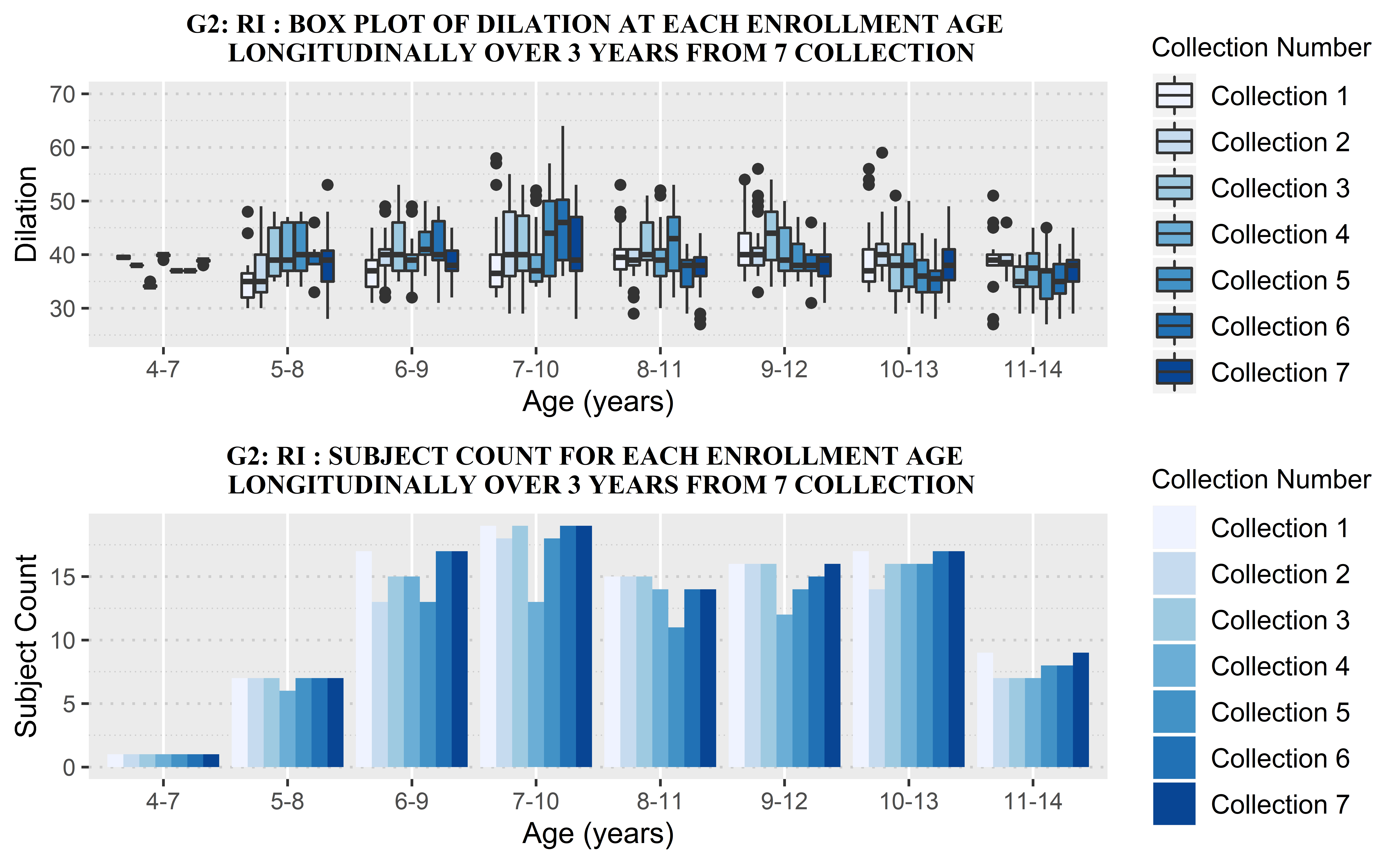}
  \label{fig:G3_R_aging_f_Dil}
\end{subfigure}

\caption{Dilation as a factor of aging for each enrollment age, over three years from seven collections and the number of subjects analyzed for each boxplot for G2 for both left and right iris (top and bottom respectively)}
\label{fig:Aging_Dil_G2}
\end{figure*}

\begin{figure}[!ht]
\centering
\begin{subfigure}{0.5\textwidth}
  \centering
  \includegraphics[width=1\linewidth, height = 4.6 cm]{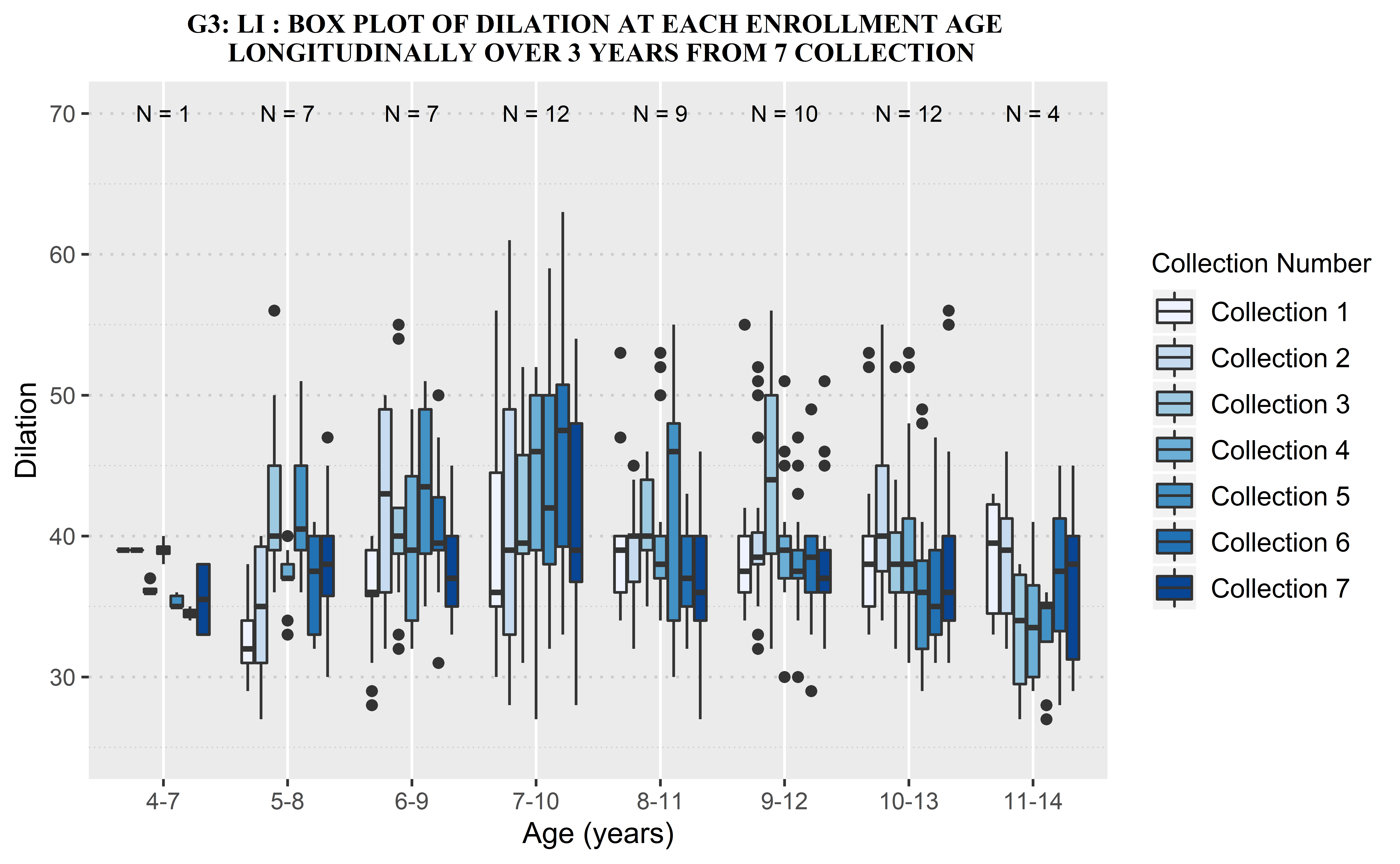}
  
  \label{fig:G3_L_aging_f_Dil}
\end{subfigure}

\begin{subfigure}{0.5\textwidth}
  \centering
  \includegraphics[width=1\linewidth, height = 4.6 cm]{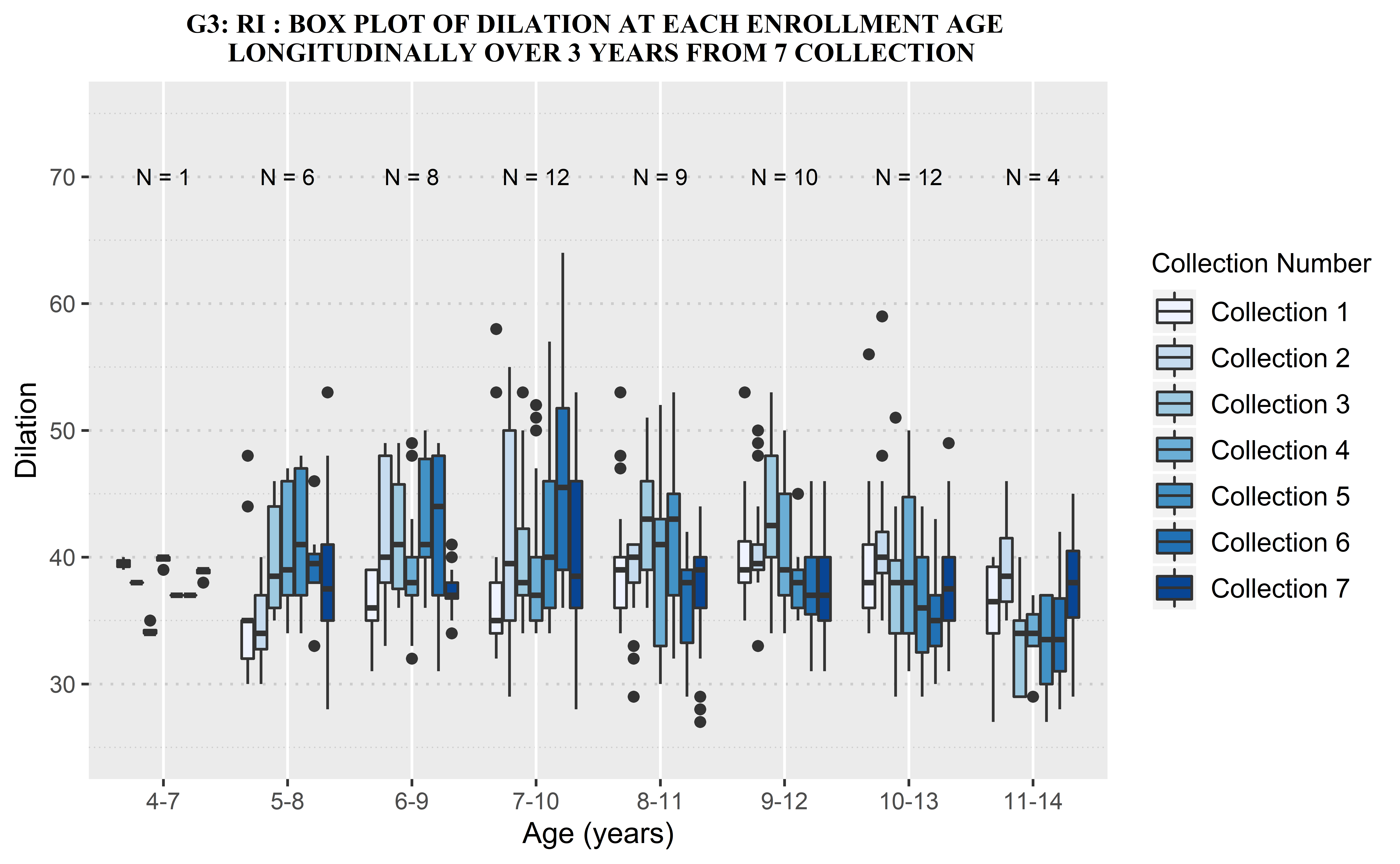}

\label{fig:Aging_Dil}
\end{subfigure}

\caption{Dilation as a factor of aging for each enrollment age, over three years from seven sessions for G3 for both LI and RI (top and bottom respectively). `N' denotes subject count}
\label{fig:Aging_Dil_G3}
\end{figure}

 This section performs longitudinal analysis (refer Section~\ref{Lstudy}) i.e. change in dilation over time. Figure~\ref{fig:Aging_Dil_G2} and Figure~\ref{fig:Aging_Dil_G3} shows boxplots of the variation in dilation over seven sessions for three years for each enrollment age between four and 11 years for groups G2 and G3 respectively and the subject count associated with each plot. G1 is eliminated from this analysis to simplify the segregation of data and its analysis. It is important to note that the subject count in each age group for G3 is low. Two important points were observed in this analysis: (1) The variability in dilation is substantially high over all 3 years for the enrollment age group of 7 years through 10 years in both G2 ad G3 for both LI and RI. This observation is in line with our finding in Section~\ref{A_F_D} that high variability is observed in the age group of 7-10 years. However, this conclusion is not strongly reflected in the age group of 6-9 years and 8-10 years, where inconsistent variability is noted. Thus, this observation remains inconclusive. (2) Considering the dilation value at 45 as the limit, the number of outliers increases as the enrollment age increases for both groups- G2 and G3. 
 
\subsection{Aging, Delta Dilation and Match Score}
\begin{figure*}[!ht]
 \centering
 \includegraphics[width=1\textwidth, height= 9 cm]{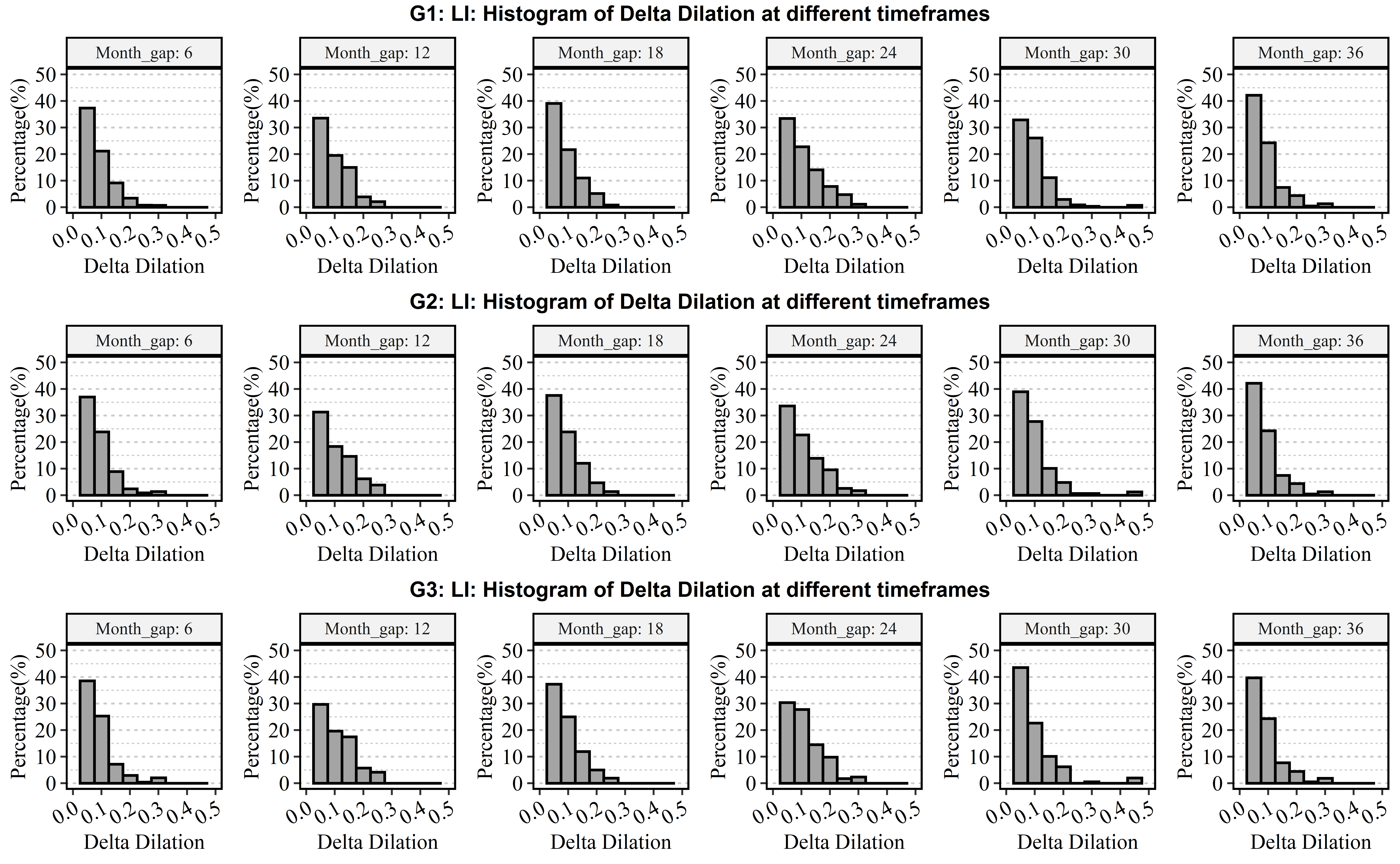}
 
\caption{Histogram of Delta Dilation, with bin width of 0.05, for different time frames of 6 to 36 months (left to right) longitudinally over 3 years for G1, G2 and G3 (top to bottom) for  left iris. RI histogram is similar to that of LI and thus is not included to accommodate space}
\label{fig:L_hist_DD}
\end{figure*}

Delta dilation in terms of longitudinal data reflects the difference in dilation between two mated pairs of images captured at different time instances and this section analyzes its impact on the match score, quantifying the performance of iris recognition in a longitudinal scenario. In our dataset $\Delta$D varies between 0 to 0.45 (0 to 45\%). The range of variation is high. Thus we looked at the distribution of $\Delta$D at different time frames for all three groups of study. Figure~\ref{fig:L_hist_DD} shows the histogram of $\Delta$D with a bin size of 0.05 at different time frames for all three groups of study for left iris. The distribution is found to be consistent across all domain (time-frame, groups of study and left and right iris). This observation supports consistency in our data collection. The majority of variation in $\Delta$D is between 5\% to 15\% (0.05 to 0.15) with a small percentage of outliers at a maximum of 45\% (0.45) $\Delta$D. We conclude that in average 10\% to 15\% variation in $\Delta$D could be expected in a practical scenario with factors affecting the collection like age, aging, partially controlled illumination and weather.\par
\begin{figure*}[!ht]
  \centering
    \includegraphics[width=1\textwidth, height = 9 cm]{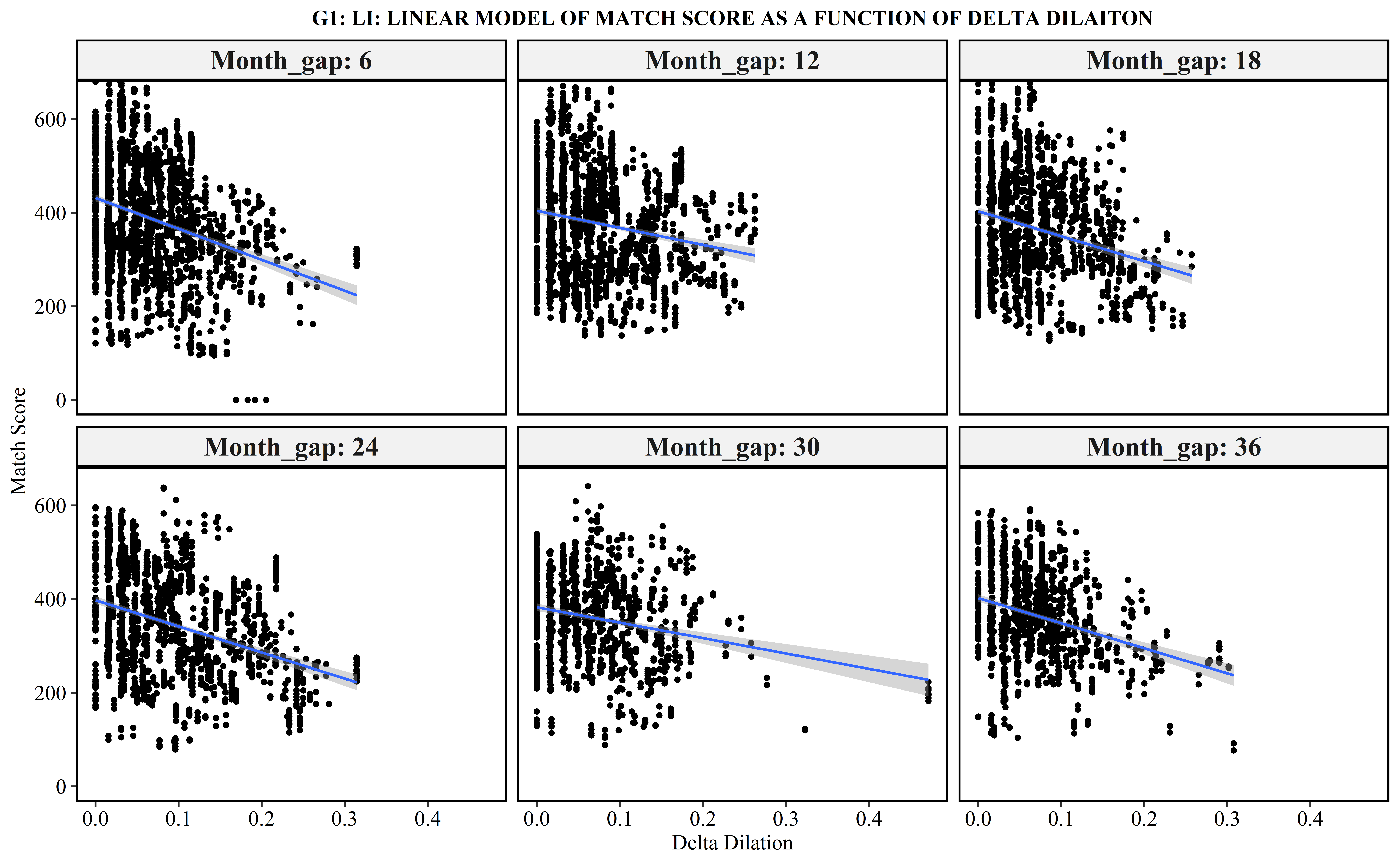}
    \caption{Linear modelling of the match score as a factor of change in dilation between enrollment and probe for each different time-frame (6 to 36 months) for G1, LI. Graphs of all other domain  (groups, left and right iris and all times frames) show similar plots, and are not included due to space constraint.}
  \label{fig:MS_DD_Aging}
\end{figure*}

   
Most previous studies indicated a linear relationship between change in dilation and any metric determining the correlation between two samples from different time frame (Match score or MS in our case). Thus we tested the linear relationship between the match score and delta dilation for our dataset by fitting a linear statistical model for different time frames of study as shown in Figure \ref{fig:MS_DD_Aging} for all three groups of studies. We note statistical significance in decay in MS with increased delta dilation for each individual time frame analysis for all groups. A statistically significant negative correlation between MS and $\Delta$D is noted across all domains (time frame, groups and left and right iris) of analysis. Overall, the estimate ranges from 328.4 $\pm$ 52.2 to 770.8  $\pm$ 44.6 with $p \geq 3.98*e^{-11} $, the slope of the model varying at different time frame. However, we do not note any trend in the decay in MS with aging (increase in time-frame from 6 month to 36 month) with the linear model. Another significant observation was the fit of the model depicted as the R-squared value, defined as the amount of variability in the data captured by the model. We note that the designed linear models only account for a minimum of 3.6\% (3.0\% in RI) to a maximum of 14.45\% (18.36\% in RI) and an average of 8.5\% (8.6\% in RI) variation in match score across all groups and all time frames due to delta dilation.\par
Assuming that $\Delta$D and MS are linearly related and the model is the best representation of the relationship, $\Delta$D induces an approximated average of 8.5\% variability in the MS. Aging is only one of many other factors affecting $\Delta$D. Thus it is safe to assume that a fraction of the 8.5\% variability is induced due to aging. Thus, only a small percentage of variability in MS could be deduced due to dilation effected as a factor of aging in children in the age group of 4 to 11 years at enrollment over a period of three years.\par
 No case of false rejection is noted due to high $\Delta$D. Over  three years, two cases of false rejection has been noted as below-
 \begin{itemize}[leftmargin=*, noitemsep]
     \item Case 1 - Left iris images rejected at 6 month time frame; the average $\Delta$D of the rejected images being 0.0901 
     \item Case 2 - Right iris images rejected at 18 month and 30 month time frame; the average $\Delta$D of the rejected images being 0.1011 and 0.105 respectively
 \end{itemize}
 In both the cases the average $\Delta$D is comfortably below the average $\Delta$D in the population. And as such, the false rejection do not appear to be the result of difference in dilation between the gallery and probe; the causes for the false rejections are beyond the scope of this paper. 
 
\section{Discussion and Conclusion}
In an attempt to understand and quantify the impact of age on dilation in children we conclude from our study that the dilation is minimum at age 4 and gradually increases and reaches it's maximum at around 8-10 years age and then gradually decreases till age 14. We conclude, on average $\Delta$D varies between 10\% to 15\% irrespective of time frame i.e. aging over a period of three years. No trend is noted in the variation in $\Delta$D with increasing time frame. No false rejection was noted due to $\Delta$D. Based on our linear model, on average $\Delta$D accounts for only 8.5\% of variability of MS. Thus we conclude that aging impacts the match score by a fraction of 8.5\%. This study cannot conclude anything beyond the studied age group of four to 14 years. Though dilation may vary across subjects and ages, we conclude that in a time period of three years the impact of aging on iris recognition performance is negligible for the age group of 4 years to 11 years.\par
It is extremely challenging to segregate the impact of different factors- age, aging, illumination, weather and medical factors on dilation. There is no publicly available dataset in this age group of four to 14 years concentrating specifically on the impact of age and aging on dilation and delta dilation for research. The data used for this study is collected from 209 individuals in the age group of four to 11 years over a period of three years spaced by approximately six months. Measures has been adapted to collect data in a partially controlled environment to minimize the impact of the variable environmental factors; however it is not completely void of the other variability factors affecting dilation.\par
 Any data that identifies an individual is sensitive, more so when it is biometric data of children. To protect the privacy of the data and the subjects, the dataset is not publicly available presently. However, we understand the lack of data and the need for such datasets to advance research in biometrics concerning children. In view of this, we are making efforts to provide access to the dataset for algorithm testing for research purposes while protecting the privacy of the data and the subjects. Such efforts needs substantial resources and time. We plan to make it available by December 2020. \par
 This paper analyzed dilation changes in children for different ages and over time. State of the art software, VeriEye, a commercially available ISO standardized \cite{iris_standard_report} software, was used for locating the pupil and outside of the iris in order to measure dilation and delta dilation. These measures are independent of the iris matching algorithm used. VeriEye was also used for iris matching. The effect of dilation on iris matching performance may be impacted by the choice of algorithm and its robustness to variation in dilation. Many techniques for iris recognition has been proposed in the literature in the last three decades \cite{daugman1993high} \cite{wildes1996machine} \cite{tisse2002person} \cite{ma2002iris} \cite{ma2004efficient} \cite{monro2007dct} \cite{sun2008ordinal} \cite{daugman2009iris} \cite{liu2016deepiris} \cite{gangwar2016deepirisnet} \cite{nguyen2017iris} \cite{arsalan2017deep}. Of these, Daugman's iris recognition algorithm \cite{daugman1993high} is the oldest and is widely adapted. However, like most commercial systems, the algorithm used by VeriEye is a blackbox. There is no public information on what measures (if any) are taken to address the impact of dilation on recognition performance or what feature extraction and matching techniques are incorporated in the algorithm. However, the results of this study are useful for practical applications which would likely to use a commercial algorithm like Verieye. We intend to study the performance of the dataset with additional algorithms, as well as allow organizations to upload their own algorithms to test against this dataset in the near future.
 
\section*{Acknowledgement}
We extend our gratitude to the Potsdam Elementary and Middle School, the staff, the participants and the parents of the participants for helping us to successfully create the invaluable dataset for this study. We also thank all the collectors from our research team and other associates for their contribution with their invaluable time. This material is based upon work supported by the Center for Identification Technology Research and the National Science Foundation under Grant No. Clarkson 1650503.

{\small
\bibliographystyle{ieee}
\bibliography{bibliography}
}

\end{document}